\documentclass[aps,floatfix,showpacs,nofootinbib,,amsmath,amssymb]
{revtex4}
\usepackage[dvips]{graphicx}

\begin{document}

\title{Light-cone observations and cosmological models: implications for
inhomogeneous models mimicking dark energy}

\author{Edward W.\ Kolb} \email{Rocky.Kolb@uchicago.edu} 

\author{Callum R.\ Lamb} \email{callum.r.lamb@gmail.com}

\affiliation{Department of Astronomy and Astrophysics, Enrico Fermi
Institute, and  Kavli Institute for Cosmological Physics, the University of
Chicago, Chicago, Illinois \ \ 60637-1433 }

\begin{abstract}

Cosmological observables are used to construct cosmological models.  Since
cosmological observations are limited to the light cone, a fixed number of
observables (even measured to arbitrary accuracy)  may not uniquely determine a
cosmological model without additional assumptions or considerations.  A
prescription for constructing a spherically symmetric, inhomogeneous
cosmological model that exactly reproduces the luminosity-distance as a function
of redshift and the light-cone mass density as a function of redshift of a
$\Lambda$CDM model is employed to gain insight into how an inhomogeneous
cosmological model might mimic dark energy models.

\end{abstract}

\pacs{98.70.Cq}

\maketitle

\section{Introduction \label{intro}}

In the last decade or so remarkable progress has been made in measuring
cosmological parameters to unprecedented accuracy. Parameters such as the Hubble
constant, $H_0=72\pm 8 \textrm{ km s}^{-1}\textrm{ Mpc}^{-1}$ \cite{Hubblekey},
the temperature of the cosmic background radiation (CBR), $T_0=2.728\pm 0.004\
\textrm{K}$ \cite{COBET}, and many other parameters, are now known to impressive
precision.  Of course, we don't invest so much time and effort determining
cosmological parameters because of an interest in numerology, but rather,
because the parameters are necessary inputs for the task of constructing a
standard cosmological model.

In turn, we do not construct cosmological models simply to make predictions to
compare with observations, but rather, to guide us to, in the words of Einstein
\cite{Einstein}, ``a deeper and more consistent comprehension of the physical
and astronomical facts.''

The precision cosmological measurements have lead to the latest cosmological
model, usually called the \textit{standard cosmological model,} the
\textit{concordance model}, or simply $\Lambda$CDM, where $\Lambda$ indicates
the inclusion of  Einstein's cosmological constant (or more generally, dark
energy), and CDM stands for cold dark matter. The model is but one possible
realization of homogeneous and isotropic cosmological solutions to Einstein's
equations. (We will refer to homogeneous/isotropic models as
Friedmann--Lema\"{i}tre--Robertson--Walker (FLRW) models.)  The question yet to
be answered is whether today's standard model will lead us to a deeper and more
consistent comprehension of the physical and astronomical facts.

A troubling feature of the $\Lambda$CDM model is the present composition of the
universe: $95\%$ is dark!  Of the dark components, roughly $25\%$ of the total
mass-energy density is dark matter, associated with galaxies, clusters of
galaxies, and other bound structures.  The bulk of the mass-energy of the
universe, about $70\%$ in the standard $\Lambda$CDM model, is attributed to dark
energy, which is usually said to drive an accelerated expansion of the Universe.

The observations and phenomena that lead to the assumptions of dark matter and
dark energy are real, but that does not necessarily imply that dark matter and
dark energy as usually envisioned are real.  Dark matter and dark energy are
good names for the phenomena.  But naming is not explaining!

In some sense, an even more remarkable feature of the standard cosmological
model is that it seems capable of accounting for all cosmological observations;
\textit{i.e.,} it seems to work!  Here, we address the issue of whether the
agreement with observations prejudices our judgment regarding the likelihood 
that the $\Lambda$CDM model is the final answer. In particular, we will discuss
the possibility that dark energy, \textit{per se}, does not exist.

The most important point regarding dark energy (and also, for that matter, the
acceleration of the expansion of the universe), which is central to the
motivation for this paper, is that all evidence for dark energy is
\textit{indirect.} The only effect of dark energy is on the expansion history of
the Universe.

The expansion rate of the Universe is perhaps the most fundamental quantity in
cosmology.  In the standard cosmological model the expansion rate is determined
from the $00$-component of the Einstein equations, $G_{00}=\kappa T_{00}$, where
$\kappa = 8\pi G$.  In FLRW models, $G_{00}=3H^2 + k/a^2$, where $H=\dot{a}/a$
is the expansion rate (here, $a$ is the Robertson-Walker scale factor and
$k=\pm1$ or $0$, depending on the geometry).  With the assumption of a
perfect-fluid stress tensor, $T_{00}=\rho$, where $\rho$ is the mass density.

The stress tensor is assumed to consist of several components with different
equations of state.  For any component $i$, the equation of state for that
component is defined as $w_i=p_i/\rho_i$.  If $w_i$ is constant, the energy
density in component $i$ evolves with redshift $z\equiv a_0/a-1$  as $\rho_i(z)
= \rho_i(0)\left(1+z\right)^{3(1+w_i)}$, where $a_0$ is the present value of the
scale factor.   Then the expansion rate as a function of redshift can be
expressed as
\begin{equation}
H^2(z)=H_0^2\left\{
\Omega_\Lambda \exp\left[\int_0^z\frac{dz_1}{1+z_1} 3 
\left[1+w_\Lambda(z_1)\right] \right]
+ \Omega_k(1+z)^2 + \Omega_M(1+z)^3 + \Omega_R(1+z)^4  \right\} .
\label{hsquared}
\end{equation}
Here, $\Omega_i$ is the ratio of the present energy density in component $i$
compared to the critical density $\rho_C=3H_0^2/\kappa$.  The subscript ``$k$''
indicates a spatial-curvature contribution ($w_k=-1/3)$, ``$M$'' indicates a
matter component ($w_M=0$), ``$R$'' indicates a radiation component
($w_R=1/3$),  and ``$\Lambda$'' represents dark energy.    In the event that the
function  $w_\Lambda(z)$ in the last term is independent of $z$, the last term
in $H^2(z)$ becomes $\Omega_\Lambda(1+z)^{3(1+w_\Lambda)}$.  The equation of
state for a cosmological constant is $w_\Lambda=-1$.

It appears that the expansion rate as a function of $z$ depends on four
constants $\{\Omega_k,\Omega_{M},\Omega_R,\Omega_\Lambda\}$ and one function
$w_\Lambda(z)$.  In reality, the situation is much more predictive.  First,
since $H^2(0)=H_0^2$, it follows that $\sum_i\Omega_i=1$, removing one
constant. 

The value of $\Omega_k$ is well determined by WMAP to be small,
$\Omega_k=-0.0026^{+0.0066}_{-0.0064}$ at the $68\%$ confidence level
\cite{wmap5}, effectively removing yet another constant. The value of $\Omega_R$
is also well determined by CBR measurements ($5\times 10^{-5}$) removing a
further constant.  One may then make the less secure assumption that $\Omega_M$
is ``known,'' or more profitably, use observational constraints to marginalize
over the $\Omega_i$'s, leaving only the function $w_\Lambda(z)$ undetermined. 
The function must be determined through its effect on the expansion history of
the Universe, which is manifest through various cosmological observables.

Many cosmological observables depend on the expansion history of the
Universe.  They often depend on the expansion history through the
coordinate distance $r$ of a source of redshift $z$.  Recall that the
Robertson-Walker metric can be written in the form
\begin{equation}
ds^2 = dt^2 - a^2(t)\left[\frac{dr^2}{1-kr^2}+r^2d\Omega^2\right] ,
\label{rw}
\end{equation}
where here $d\Omega$ is the angular differential and $r$ is the
comoving ``radial'' coordinate.  The radial coordinate of a source at
redshift $z$ is determined by integrating the null geodesic equation
($ds^2=0$) to obtain
\begin{equation}
r(z) = \left. \begin{array}{l} \sin \\ 1 \\ \sinh \end{array} \right\}
\left[ \int_0^z \frac{dz'}{H(z')} \right] ,
\label{rofz}
\end{equation}   
where $\sin$, $1$, $\sinh$ obtains for $k=+1,\ 0, \-1$, respectively.

Observables such as the luminosity distance, the angular-diameter distance, the
volume element, and the age of the Universe, all depend on the time evolution of
the expansion rate, hence on the properties of the stress tensor, hence on the
dark energy fluid characterized by $w_\Lambda$.

At present, a constant $w_\Lambda(z)$ seems to fit the data, and furthermore
$w_\Lambda=-1$, the value of $w_\Lambda$ for a cosmological constant, is not
disallowed.  It is also true that since there is no apparent reason to exclude
vacuum energy from the stress tensor, there is no reason \textit{not} to imagine
it enters the determination of the expansion rate.  The problem is that the
magnitude of the  expected vacuum energy is very many orders of magnitude larger
than expected. If the cosmological constant is part of nature, then either some
principle must be found to explain the much smaller-than-expected magnitude of
$\Omega_\Lambda$, or one must resort to anthropic arguments.

This situation has led many to consider alternate explanations of the
observations: modified gravity, new long-range forces, new ultra-light scalar
fields, extra dimensions, branes, bulk, Lorentz-invariance violation,
\textit{etc.}   Some of these approaches are, in many respects, more drastic
than a small cosmological constant.  For instance, theories of modified gravity
typically do not respect the principle of general covariance.  Since we should
not abandon our principles without good cause, it is tempting to retreat to the
position that since a cosmological constant is adequate to account for the data,
just assume that it exists and call it a day?

We resist the retreat because our goal in cosmology is more than just explaining
the observations.  The ingredients of a cosmological model must be deeply
grounded in fundamental physics.  Dark matter, dark energy, modified gravity,
mysterious new forces and particles, {\em etc.,} unless part of an overarching
model of nature, should not be part of a cosmological model.  We may plant new
ideas and features into cosmological models, but they will ultimately prove
barren unless grounded in fundamental physics.  No matter how well a concordance
model fits the data, its ingredients must not be discordant with fundamental
physics.

In this paper we explore a less traveled path, and consider the possibility that
the Friedmann equation is not an adequate description of our inhomogeneous
Universe  
\cite{Celerier:1999hp,Buchert:1999mc,Schwarz:2002ba,Rasanen:2003fy,Kolb:2004am,Rasanen:2004js,Kolb:2005me,Rasanen:2005zy,Barausse:2005nf,Notari:2005xk,Rasanen:2006kp}. 
In this approach there is no need for dark energy, so no need for new long-range
forces, modifications of general relativity, new ultralight particles, or
anthropic reasoning.  The phenomenon of inhomogeneities mimicking dark energy
is sometimes referred to as ``backreaction.''

In this paper we follow the prescription of Mustapha, Hellaby, and Ellis
\cite{Mustapha:1998jb} [MHE], recently applied to $\Lambda$CDM models by
C\'{e}l\'{e}rier, Bolejko, Krasinski, and Hellaby
\cite{Celerier:2009sv},\footnote{Similar formalism was developed by Chung and
Romano \cite{Chung:2006xh} to fit only the luminosity-distance, and by Yoo, Kai,
and Nakao \cite{Yoo:2008su} who simultaneous fit the luminosity distance and
require no spatial variations of the age of the Universe.} and explore the
implications of the fact that one can construct a spherically symmetric
inhomogeneous model that \textit{exactly} reproduces the redshift dependence of
the luminosity distance and mass density of any given $\Lambda$CDM model.  We
show that in general one expects the cosmological solution thus obtained to have
a ``mixmaster'' approach to the initial singularity, with the radial scale
factor diverging and the angular scale factor vanishing.  We then show that
there is another cosmological solution with the same density and curvature
functions as the singular model  but with regular initial conditions where both
scale factors vanish at the singularity, and that this regular solution well
\textit{approximates} another $\Lambda$CDM model.  We then discuss the utility
of different averaging prescriptions for understanding the results.

In the next section we briefly review spherically symmetric cosmological
solutions.  In Sect.\ \ref{recon} we review the construction of such an
inhomogeneous model that reproduces key observational features of $\Lambda$CDM.
In Sect.\ \ref{bang} we evolve the model off the light cone, paying special
attention to its behavior near the bang.  In Sect.\ \ref{average} we calculate
some spatially averaged features of the model.  Section \ref{conclusions}
presents the conclusions and discusses their implications.

\section{Lema\"{i}tre--Tolman--Bondi Models \label{LTB}}

In this  section we discuss Lema\"{i}tre--Tolman--Bondi
\cite{Lemaitre:1933gd,Tolman:1934za,Bondi:1947av}  (LTB) models in general, and
in the next section we discuss a particular LTB model that \textit{exactly}
reproduces select observational features of the $\Lambda$CDM model.

LTB models are spherically symmetric cosmological solutions to the Einstein
equations with a dust stress-energy tensor.   The LTB model is based on the
assumptions that the system has purely radial motion and the motion is geodesic
without shell crossing (otherwise the pressure could not be neglected). The 
line element in the synchronous gauge can be written in the form
\begin{equation}
\label{ltbmetric}
ds^2=-dt^2 + \frac{R'^2(r,t)}{1+\beta(r)} dr^2 + R^2(r,t) \, d\Omega^2 .
\end{equation}
Here, the prime superscript denotes $d/dr$.  We will denote $d/dt$ by an
overdot.  The function $\beta(r)$ is an arbitrary function of $r$.  The
Robertson--Walker metric can be recovered if $R(r,t)\rightarrow a(t)r$ and
$\beta(r)\rightarrow-kr^2$.

In spherically symmetric models, in general there are two expansion rates: an
angular expansion rate, $H_\perp\equiv \dot{R}(r,t)/R(r,t)$, and a radial
expansion rate, $H_r\equiv \dot{R}'(r,t)/R'(r,t)$. With the dust equation of 
state, the Einstein equations may be expressed as 
\begin{eqnarray}
H_\perp^2(r,t) + 2H_r(r,t)H_\perp(r,t) -\frac{\beta(r)}{R^2(r,t)} 
- \frac{\beta'(r)}{R(r,t)R'(r,t)} & = & \kappa\rho(r,t)
\nonumber \\
6\frac{\ddot{R}(r,t)}{R(r,t)} + 2H_\perp^2(r,t) - 2\frac{\beta(r)}{R^2(r,t)} 
- 2H_r(r,t)H_\perp(r,t) + \frac{\beta'(r)}{R(r,t)R'(r,t)} 
& = & -\kappa\rho(r,t) .
\end{eqnarray}
These represent the generalization of the Friedmann equation for a
homogeneous/isotropic universe to a spherically symmetric inhomogeneous
universe.

The equations may be manipulated to result in an easily integrable dynamical 
equation for $R(r,t)$:
\begin{equation}
\label{rdot}
\dot{R}(r,t)=\sqrt{\beta(r)+\frac{\alpha(r)}{R(r,t)}} ,
\end{equation}
where $\alpha(r)$ is a function of $r$ related to the density $\rho(r,t)$ by
\begin{equation}
\kappa\rho(r,t)=\frac{\alpha'(r)}{R^2(r,t)R'(r,t)} .
\end{equation}
Equation (\ref{rdot}) can be differentiated to yield the dynamical equation 
for $R'(r,t)$:
\begin{equation}
\label{rpdot}
\dot{R}'(r,t) = \frac{\beta'(r)+\alpha'(r)/R(r,t)-\alpha(r) R'(r,t)/R^2(r,t)}
{2\dot{R}(r,t)} .
\end{equation}

The two functions $\alpha(r)$ and $\beta(r)$ define the LTB model, playing roles
analogous to the total energy density parameter $\Omega_0$ and the spatial
curvature parameter $\Omega_k$ of FLRW models. To see this, recall that the
evolution of the scale  factor and time for a hyperbolic dust Friedmann model
evolves as ($\eta$ is conformal time)
\begin{equation}
a(t) = a_0 \frac{\Omega_0}{2\Omega_k}
	\left[\cosh\eta -  1   \right] ,
\qquad		
t-t_{BB} = H_0^{-1}\frac {\Omega_0}{2\Omega_k^{3/2}}
	\left[\sinh\eta - \eta \right] ,
\end{equation}
where $a_0$ is the present value of the scale factor and $t_{BB}$ is the time of
the big bang (usually set to zero).

Since Friedmann models are homogeneous and isotropic, they are zero-dimensional
models.\footnote{The dimension of a model refers to the number of dynamically
independent combinations of spatial dimensions that enter in physical
quantities.}  The LTB models are spherically symmetric, so they are
one-dimensional problems.  So for LTB models the density parameter $\Omega_0$ is
replaced by a density function $\alpha(r)$ and the curvature parameter
$\Omega_k$ is replaced by a curvature function $\beta(r)$.  If $\beta(r) > 0$,
which is the analog of hyperbolic Friedmann models, the evolution of the metric
function $R(r,t)$ can be described by
\begin{equation}
\label{rt}
R(r,t) = \frac{\alpha(r)}{2\beta(r)}
	\left[\cosh\eta(r) - 1 \right] ;
\qquad
t-t_{BB}(r) = \frac{\alpha(r)}{2\beta^{3/2}(r)}
	\left[\sinh\eta(r) - \eta(r)\right],
\end{equation}
where now, in general, the bang time, $t_{BB}(r)$ is a function of $r$, as is
the effective conformal time $\eta(r)$.

The photon geodesic equation for the position of the photon as a function of 
time, $\hat{t}(r)$, is found from the LTB metric:
\begin{equation}
\label{tofr}
\frac{d\hat{t}(r)}{dr}=-\frac{R'(r,\hat{t}(r))}{\sqrt{1+\beta(r)}}. 
\end{equation}
The redshift of the photon, $z(r)$, is
\begin{equation}
\label{zofr}
\frac{dz(r)}{dr}=(1+z)\frac{\dot{R}'(r, \hat{t}(r))}{\sqrt{1+\beta(r)}}.
\end{equation}
Equations (\ref{tofr}) and (\ref{zofr}) are solved with initial conditions
$z(r=0)=0$ and $\hat{t}(r=0)=0$. The luminosity distance is then simply
\begin{equation}
\label{dlz}
\hat{d}_L(z)=(1+z)^2R(r,\hat{t}(r)).
\end{equation} 
(We denote the luminosity distance by $\hat{d}_L$ to emphasize that it is a
light-cone observable.)

\section{Reconstructing an LTB model with $\Lambda$CDM observational features
\label{recon}}

In this section we will review the procedure and results that allow one to
construct an LTB model that reproduces 
\begin{enumerate}
\item the luminosity-distance--redshift relationship, $\hat{d}_L(z)$, of a 
fiducial $\Lambda$CDM model, and
\item the light-cone matter density as a function of redshift, 
$\widehat{\rho}(z)$, of the fiducial $\Lambda$CDM model.
\end{enumerate}

The ``fiducial'' $\Lambda$CDM model may in principle be any FLRW model with a
cosmological constant.  In this paper we will assume for the fiducial model a
spatially-flat Universe, so $\Omega_M +\Omega_\Lambda=1$.  Unless otherwise
specified, we will also assume $\Omega_\Lambda=0.7$ (so $\Omega_M=0.3$).  The
assumption of spatial flatness will simplify our calculations.  Neither that
assumption, nor the exact value of $\Omega_\Lambda$, is important in our
considerations.

The reconstruction method we follow was discussed in general by Mustapha,
Hellaby, and Ellis in 1998 \cite{Mustapha:1998jb} [MHE], and recently applied to
$\Lambda$CDM models by C\'{e}l\'{e}rier, Bolejko, Krasinski, and Hellaby [CBKH],
\cite{Celerier:2009sv}. Related formalism was developed by Chung and Romano 
\cite{Chung:2006xh} and by Yoo, Kai, and Nakao \cite{Yoo:2008su}.

\begin{figure}
\begin{center}
\includegraphics[width=12cm]{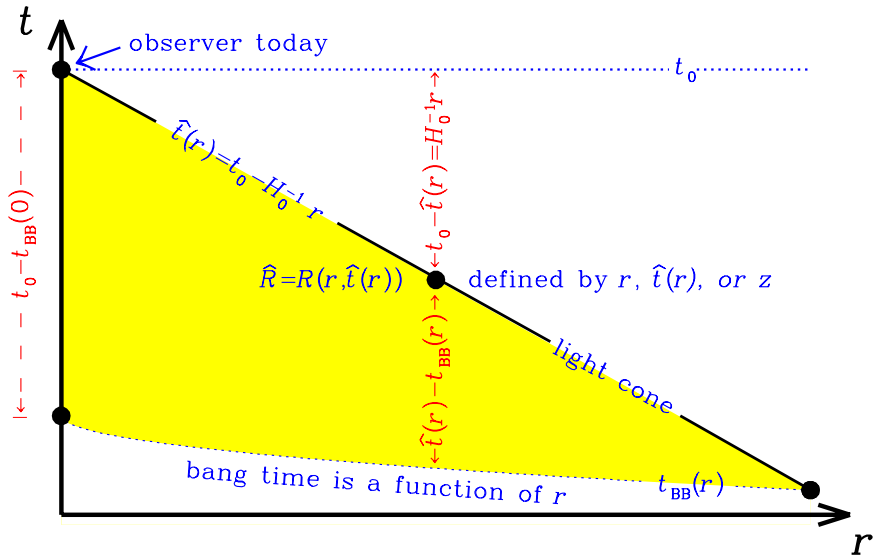}

\caption{This figure illustrates features and notation of the LTB model analyzed
in this paper. The time coordinate is defined such that its value at the present
time is $t_0$; $t_0$ is not the present {\em age} of the Universe. In LTB models
the age of the universe is a function of $r$, since, in general, the bang time,
$t_{BB}$, is a function of $r$.  The age as a function of $r$ is given by
$t_0-t_{BB}(r)$.  The light cone (the past-directed null geodesic) for an
observer at $r=0$ is denoted by the solid line.  A ``hat'' indicates that a
quantity is evaluated on the light cone, and hence is a function of only one
variable: $\widehat{f}(r) \equiv f(r, \hat t(r))$. Hatted quantities are often
written as functions of $z$, since that is the light-cone coordinate accessible
to observations. The $r$ coordinate is scaled such that $\hat{t}(r)$ is given by
$\hat{t}(r)=t_0-H_0^{-1}r$.  The values of $R$ and $R^\prime$ on the light cone
are denoted by $\widehat{R}=R(r,\hat{t}(r))$ and
$\widehat{R}^\prime=R^\prime(r,\hat{t}(r))$.  The shaded area is our causal
past.}

\label{rhat}

\end{center} 
\end{figure}

Let us review the procedure.  To follow the discussion, it is useful to refer to
Fig.\ \ref{rhat}.  Observers obtain information about the universe by means of
photons, so our direct information about the Universe is limited to the light
cone given by  Eq.\ (\ref{tofr}).  Quantities evaluated on the light cone are
denoted by a  ``hat.''  Therefore, the metric functions $R(r,t)$ and
$R^\prime(r,t)$, and the density $\rho(r,t)$ evaluated on the light cone, are
\begin{equation}
R(r,\hat{t}(r)) \equiv \widehat{R} ; \qquad
R^\prime(r,\hat{t}(r)) \equiv \widehat{R}^\prime; \qquad
\rho(r,\hat{t}(r)) \equiv \widehat{\rho}.
\end{equation}

The light cone for an observer situated at the origin is illustrated in Fig.\
\ref{rhat} by the solid line.  The shaded area indicates the causal volume of
the Universe within the light cone.  Again, we emphasize that \textit{all}
direct astronomical information is limited to the light cone.  We often speak of
cosmology within our causal volume, or perhaps the density field at some fixed
time like the present (which includes an acausal region of the Universe). We do
this even though we have no direct observational evidence for it.  To speak of
the Universe in this way we must employ a cosmological model that allows us to
depart from the light cone (in theory, not literally).  What we say about the
Universe off the light cone is only as trustworthy as the cosmological model
used to generate the information.  But as emphasized by MHE, light-cone
observations (or at least the two light-cone observations we use) do not
uniquely determine the cosmological model on the light-cone (\textit{i.e.}, both
the $\Lambda$CDM and LTB models give the same results), so of course they do not
uniquely determine the cosmological model off of the light cone.

As in MHE, we take advantage of a coordinate freedom to simplify the
calculation.  The radial coordinate $r$ has no physical significance; it can be
rescaled in any convenient way. Here, we rescale $r$ such that \textit{on the
light cone}
\begin{equation}
\widehat{R}^\prime=H_0^{-1}\sqrt{1+\beta(r)}.  
\end{equation}

\begin{figure}
\begin{center}
\includegraphics[width=12cm]{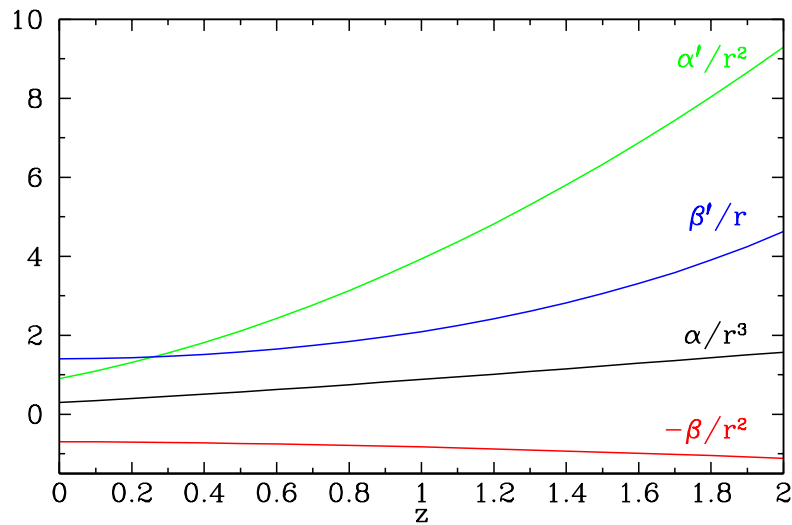}

\caption{The functions $\alpha(z)$, $\beta(z)$, $\alpha^\prime(z)$, and
$\beta^\prime(z)$ as a function of $z$.}

\label{abapbp}

\end{center} 
\end{figure}

In order to find $\alpha(r)$ and $\beta(r)$ we follow the procedure outlined by
MHE and followed in CBKH.  Here we take the view that we will construct an LTB
model (\textit{i.e.,} find $\alpha(r)$ and $\beta(r)$) that agrees with
observations of $\hat{d}_L(z)$ and $\widehat{\rho}(z)$ only out to $r$
corresponding to $z=2$, and beyond that we have no useful information about
$\hat{d}_L(z)$ or $\widehat{\rho}(z)$ that would constrain  $\alpha(r)$ and
$\beta(r)$.  Perhaps information from cosmic microwave background (CMB)
measurements could fill that niche, but since the purpose of this paper is to
investigate constraints from the local Universe, we will not pursue constructing
the model beyond a value of $r$ corresponding to $z=2$.  
\begin{enumerate}
\item The first step is to use the assumption that the LTB 
luminosity-distance--redshift relationship matches that of the fiducial 
$\Lambda$CDM model.  In spatially flat $\Lambda$CDM, the luminosity distance is
given by
\begin{equation}
\hat{d}_L(z) = (1+z) \int_0^z \frac{dz_1}{H_{\Lambda\textrm{CDM}}(z_1)} ,
\end{equation}
where $H_{\Lambda\textrm{CDM}}(z)=H_0\sqrt{\Omega_M(1+z)^3+\Omega_\Lambda}$. 
Therefore, using  Eq.\ (\ref{dlz}), 
\begin{equation}
\widehat{R}(z) = \frac{H_0^{-1}}{1+z}\int_0^z
\frac{dz_1}{\sqrt{\Omega_M(1+z_1)^3+\Omega_\Lambda}}.
\end{equation}
\item The next step is enforce the assumption that the observed mass density on
the light cone in the LTB model as a function of $z$, $\widehat{\rho}(z)$,
matches the observed mass density as a function of $z$ in the fiducial 
$\Lambda$CDM model.  Using $\widehat{\rho}(z)\,
d^3V_\textrm{LTB}=\rho_{M,\Lambda\textrm{CDM}}\, d^3V_{\Lambda\textrm{CDM}}$, we
find\footnote{In MHE and CBKH, $H_0^{-1} \widehat{R}^2(z) \kappa
\widehat{\rho}_\textrm{LTB}(z) dr  = \kappa \widehat{m}ndz$.}
\begin{equation}
H_0^{-1} \widehat{R}^2(z)\, \kappa\, \widehat{\rho}(z)\, \frac{dr}{dz} = 
 \Omega_M 3 \frac{H_0^2}{H_{\Lambda\textrm{CDM}}(z)} 
\left[ \int_0^z\frac{dz_1}{H_{\Lambda\textrm{CDM}}(z_1)} \right]^2 ,
\end{equation}
where we have used the fact that $\rho_{M,\Lambda\textrm{CDM}}(z) = \Omega_M
3 H_0^2 \kappa^{-1} (1+z)^3$.
\item We make use of the fact that 
\begin{equation}
\frac{dz}{dr}= \left[\frac{H_0 d\widehat{R}(z)}{dz}(1+z)\right]^{-1}
\left[ 1- \frac{1}{2}\int_0^z H_0^{-1}\kappa\widehat{\rho}(z)\widehat{R}(z_1) 
(1+z_1)
\frac{dr}{dz_1}dz_1 \right] = (1+z) \frac{H_{\Lambda\textrm{CDM}}(z)}{H_0},
\end{equation}
where the last equality (not found in MHE or CBKH) obtains only for a spatially
flat Friedmann model.  The resulting $z(r)$ is shown in Fig.\ \ref{fzofr}. 

\begin{figure}
\begin{center}
\includegraphics[width=12cm]{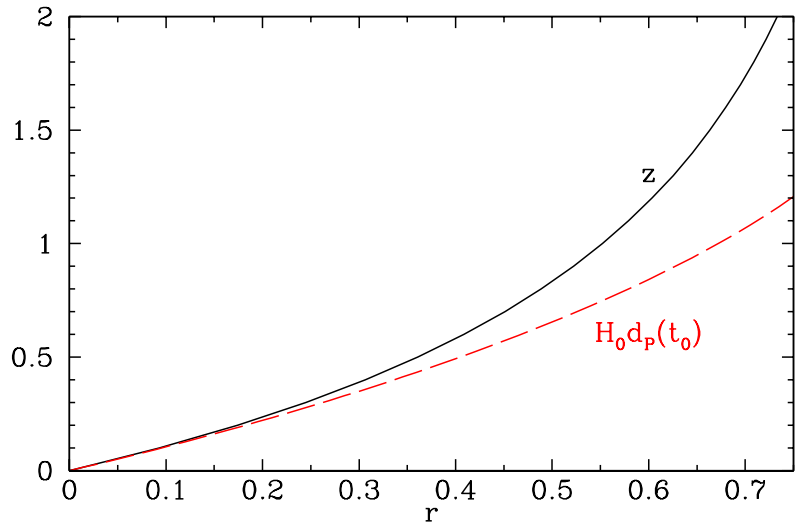}

\caption{The redshift $z$ and the present proper distance $H_0d_P(t_0)$ as a 
function of the coordinate $r$.}

\label{fzofr}

\end{center} 
\end{figure} 

\item We then solve the differential equation from MHE for $\alpha(r)$:
\begin{equation}
\label{dal}
\frac{d\alpha}{dr} = \frac{1}{2} H_0^{-1} \widehat{R}^2(z)\, \kappa\, 
\widehat{\rho}(z) \left[ \frac{1}{H_0d\widehat{R}/dr}
\left(1-\frac{\alpha}{\widehat{R}}\right) + H_0 \frac{d\widehat{R}}{dr} \right],
\end{equation}
where $d\widehat{R}/dr=d\widehat{R}/dz\cdot  dz/dr$.\footnote{Here we mention a
subtlety in Eq.\ (\ref{dal}). The factor $d\widehat{R}/dr$ vanishes at some
value of $z$ (in the fiducial $\Lambda$CDM  model, at $z\sim1.6$), but since
$1-\alpha/\widehat{R}$ also vanishes at that point, the equation is regular. 
However, care must be taken when numerically integrating the equation.}  The
initial condition is $\alpha(0)=0$.
\item Finally, $\beta(r)$ is given by
\begin{equation}
\beta(r) = \left( \frac{d\alpha}{dr} 
\frac{1}{H_0^{-1}\widehat{R}^2\kappa\widehat{\rho}}
\right)^2 -1 .
\end{equation}

The values of $\alpha$, $\beta$, $\alpha^\prime$, and $\beta^\prime$ obtained
using this procedure are shown in Fig.\ \ref{abapbp}.  They agree with the
results of CBKH.

\end{enumerate}

We emphasize that for these values of $\alpha(r)$ and $\beta(r)$, the LTB model
\textit{exactly} reproduces the luminosity-distance--redshift relation and the
mass-density--redshift relation of $\Lambda$CDM.  

Once $\alpha(r)$ and $\beta(r)$ are determined, along with $\widehat{R}$ and
$\widehat{R}^\prime$, one may use Eq.\ (\ref{rpdot}) to solve for
$R^\prime(r,t)$ and either Eq.\ (\ref{rdot}) or Eq.\ (\ref{rt}) to solve for
$R(r,t)$ from the bang time, $t_{BB}(r)$, until today, $t_0$. 

Again, we return to a point made earlier: We only have observational information
on our light cone.  This means that although one might speak of the density as a
function of $r$ at a particular time, it is not accessible to observations.
Nevertheless, it is instructive to examine $\rho(r,t)$ at fixed times.  Such an
example is the present value of $\rho$, $\rho(r,t_0)$.  CBKH present it as a
function of the metric function $R(r,t_0)$, but here, we present it in terms of
a physical quantity, namely $d_P(t_0)$, the present proper distance to $r$.  The
present proper distance is given by
\begin{equation}
H_0 d_P(t_0) = \int_0^r dr_1 \frac{R^\prime(r_1,t_0)}{\sqrt{1+\beta(r_1)}},
\end{equation}
and is shown in Fig.\ \ref{fzofr} as a function of coordinate $r$.

\begin{figure}
\begin{center}
\includegraphics[width=12cm]{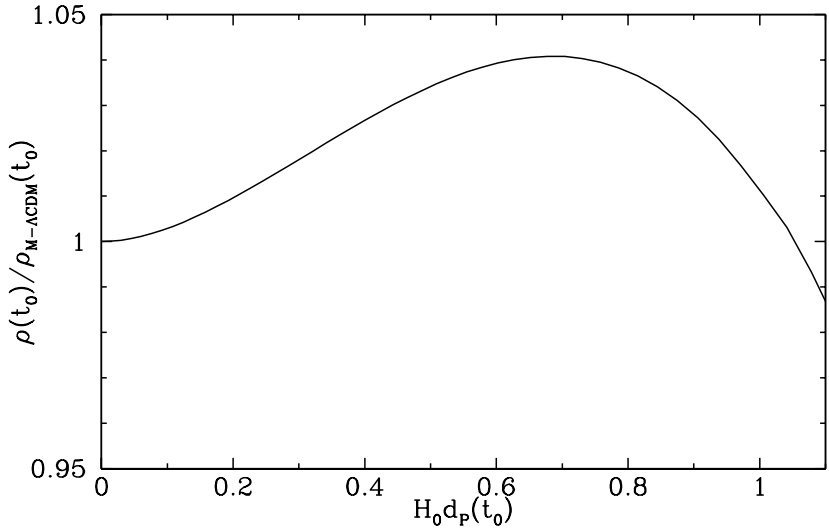}

\caption{The present value of $\rho$, $\rho(t_0)$, as a function of the present
proper distance.  It is normalized to the value in the fiducial $\Lambda$CDM
model, $\rho_{M-\Lambda\textrm{CDM}}(t_0) = 3H_0^2\kappa^{-1}\Omega_M$.}

\label{rhodp}

\end{center} 
\end{figure} 

Again, as noted by CBKH, while previous attempts to construct an LTB model to
reproduce $\hat{d}_L(z)$ of the fiducial $\Lambda$CDM model have resorted to a
large local void
\cite{Celerier:1999hp,Tomita,Iguchi:2001sq,Moffat,Alnes:2005rw,Mansouri:2005rf,Vanderveld:2006rb,Garfinkle:2006sb,Biswas:2006ub,Alnes:2006uk,Caldwell:2007yu,Alexander:2007xx,GarciaBellido,Clifton:2008hv,Hunt:2008wp,Valkenburg:2009iw},
here we find a very mild \textit{increase} (less than $4\%$) in $\rho(r)$. (As
already remarked, this increase is unobservable since only $\widehat{\rho}(z)$
is observable, and that exactly agrees with $\Lambda$CDM).

A point that will be developed further is worth introducing here.  It has often
been remarked that it is not possible for inhomogeneous cosmologies to reproduce
$\hat{d}_L(z)$ of $\Lambda$CDM models because large, unobserved, inhomogeneities
would be necessary.  This is usually stated as either 1) The large
inhomogeneities would result in large, unobserved, peculiar velocities, or 2)
One can always write the metric in the perturbed Newtonian form.  These two
arguments are often said to exclude the possibility that an LTB model can mimic
dark energy observables. 

However, the present consideration seems to provide (yet another) counterexample
to that statement, since in the LTB model constructed here, $\hat{d}_L(z)$ is
exactly that of the fiducial $\Lambda$CDM model while $\widehat{\rho}(z)$ is
also exactly that of the fiducial $\Lambda$CDM model.

In the next section we explore the time evolution of the model off of the light
cone.

\section{From the bang time until today \label{bang}}

\begin{figure}
\begin{center}
\includegraphics[width=12cm]{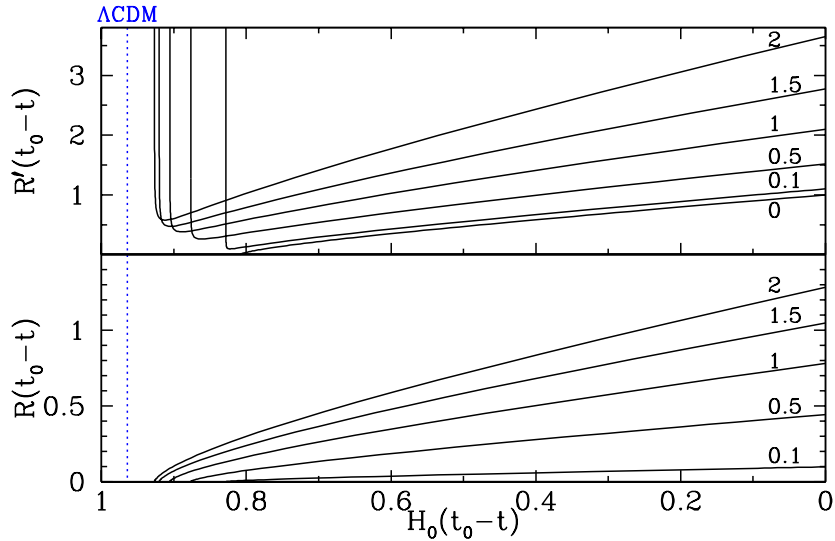}

\caption{The evolution of $R(r,t)$ and $R^\prime(r,t)$ with time for the model
discussed in Sec.\ \ref{recon}.  Labels on the curve correspond to values of $r$
corresponding to the indicated values of $z$.  $R(0,t)=0$ is not indicated.  The
age of the universe in the fiducial $\Lambda$CDM model is shown by the dotted
vertical line.}

\label{rrp}

\end{center} 
\end{figure}

Let us first examine the evolution of $R(r,t)$ and $R^\prime(r,t)$ for the model
constructed in the previous section.  Again, we only consider the evolution out
to $r$ corresponding to $z=2$.  The values of $R(r,t)$ and $R^\prime(r,t)$ for
several values of $r$ (labeled by the corresponding values of $z$) are indicated
in Fig.\ \ref{rrp}.  

The striking feature is that for $r\neq 0$, as $t_{BB}(r)$ is approached,
$R^\prime(r,t)\rightarrow \infty$.\footnote{To better appreciate why this
feature is striking, recall that for an FLRW solution, $R^\prime(r,t)
\rightarrow a(t)$, where $a$ is the scale factor.  One would expect the scale
factor to approach zero at the bang time. Here, the effective radial scale
factor $R^\prime$ blows up as the angular scale factor $R$ vanishes.}   (The
solution at the origin at the bang time is $R^\prime(0,t_{BB}(r)) =0$.)  The
behavior of the radial scale factor increasing as the angular scale factor
decreasing is reminiscent of a type of Type I Bianchi model (the so-called
Kasner solution).

To understand this behavior, consider the evolution of $R^\prime$ at early
times.  It is most convenient to work with the parameter $\eta$ rather than
time.  From Eq.\ (\ref{rt}), $dt/d\eta = R(r,t)/\beta^{1/2}$, and using Eq.\
(\ref{rpdot}), we find
\begin{equation}
\label{smalleta}
\frac{dR^\prime}{d\eta} = \frac{\alpha^\prime}{4\beta} \eta - 
\frac{R^\prime}{\eta}.
\end{equation}
Two small-$\eta$ solutions are possible.  First, if one assumes that
$R^\prime\rightarrow 0$ as $\eta\rightarrow 0$, then the small-$\eta$
solution is 
\begin{equation}
\label{regular}
R^\prime \rightarrow \frac{\alpha^\prime}{12\beta}\eta^2,
\end{equation}
which vanishes at the bang time ($\eta=0$).  We will refer to this as the
\textit{regular} solution.

The other solution to Eq.\ (\ref{smalleta}) obtains if as $\eta\rightarrow 0$,
the second term on the rhs of the equation dominates.  In this case, the
solution is
\begin{equation}
\label{singular}
R^\prime \rightarrow \frac{A}{\eta},
\end{equation}
where $A$ is an constant.  We will refer to this as the \textit{singular} 
solution.

The different behaviors near the origin can be understood by expressing
$R^\prime(r,t)$ as \cite{Hellaby:1985zz}
\begin{equation}
R^\prime(r,t) = \left( \frac{\alpha^\prime}{\alpha} - \frac{\beta^\prime}{\beta}
\right) R(r,t) - \left( \frac{\alpha^\prime}{\alpha} - \frac{3}{2}
\frac{\beta^\prime}{\beta} \right) \left[t - t_{BB}(r)\right]\dot{R}(r,t)
-t_{BB}^\prime(r) \dot{R}(r,t).
\label{rpanalytic}
\end{equation}
At $t=t_{BB}(r)$, $R(r, t_{BB}(r))=0$, and Eq.\ (\ref{rpanalytic}) implies
$R^\prime(r,t_{BB}(r)) = -t_{BB}^\prime(r) \dot{R}(r,t_{BB}(r))$.  Since
$\dot{R}(r,t_{BB}(r)) =+\infty$ and in our case $t_{BB}^\prime(r)<0$, we see
that $R^\prime(r,t_{BB}(r)) = + \infty$ as seen in Fig.\ \ref{rrp}.  Since as
$t\rightarrow 0$, 
\begin{equation}
t \rightarrow \frac{\alpha(r)}{\beta^{3/2}(r)} \frac{\eta^3(r)}{12}; \qquad
R \rightarrow \frac{\alpha(r)}{\beta(r)} \frac{\eta^2}{4}; \qquad
\dot{R} \rightarrow \frac{2\beta^{1/2}}{\eta},
\end{equation}
the constant $A$ in Eq.\ (\ref{singular}) is identified to be
$A=-t_{BB}^\prime(r) 2\beta^{1/2}$.

The solution $R^\prime(r,t_{BB}(r))=0$ only will occur if $t_{BB}^\prime(r)=0$,
or in other words, if the big-bang surface is synchronous.   In this case we
can set $t_{BB}=0$, and one can easy recover Eq.\ (\ref{regular}) from Eq.\
(\ref{rpanalytic}).

Different initial conditions, either Eq.\ (\ref{regular}) or (\ref{singular}),
will lead to different evolutions of $R^\prime(\eta)$ for the same values of
$R(\eta)$.   It is not difficult to see that the space spanned by singular
solutions is larger than the space spanned by the regular solution because the
constant $A$ (related to $t^\prime_{BB}(r)$) of the singular solutions is
undetermined. The regular initial condition leads to a \textit{unique} value of
$R^\prime$ at some later time, whereas the singular initial condition can lead
to \textit{any} value of $R^\prime$ at some later time, modulo the value
obtained from the regular initial condition.  Or as we have shown, the regular
initial condition only occurs for a simultaneous bang surface.

There is no particular reason to believe that $\widehat{R}^\prime$ would
correspond to the unique value obtained from the regular initial condition. 
This conforms to the expectation that unless there is some symmetry to prevent
divergence, the generic approach to the singularity should not be regular.

At this point it should be pointed out that the small-$\eta$ behavior might be
irrelevant for three reasons: 1) The LTB model is merely a toy model, not to be
taken seriously. 2) The LTB model we consider has only a dust component to the
stress tensor.  Adding radiation may change the outcome, or imagining an
early inflationary phase might also change the outcome. 3) Finally, perhaps it
is premature to reject the singular solution without fully exploring why it
is inappropriate, \textit{i.e.,} what is the observational basis for excluding
it?

We eschew these issues and ask a modest question of how the observations that
went into the construction of $\alpha(r)$ and $\beta(r)$ would be changed if we
substituted the regular initial conditions.  This is quite straightforward: we
have $\alpha(r)$ and $\beta(r)$, all we have to do is choose initial conditions
$R(r,0)=R^\prime(r,0)=0$, evolve the metric functions according to Eqs.\
(\ref{rdot}) and (\ref{rpdot}), enforce $t^\prime_{BB}(r)=0$, and calculate the
observables $\hat{d}_L(z)$ and $\widehat{\rho}(z)$.

\begin{figure}
\begin{center}
\includegraphics[width=12cm]{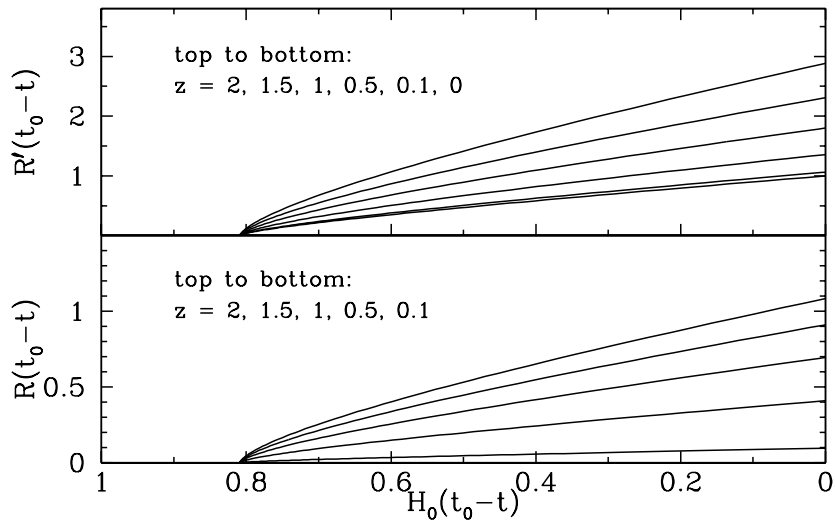}

\caption{The evolution of $R(r,t)$ and $R^\prime(r,t)$ with time for $\alpha(r)$
and $\beta(r)$ from the model discussed in Sec.\ \ref{recon}, but with the
condition that $t_{BB}^\prime(r)=0$.  Labels on the curve correspond to values
of $r$ corresponding to the indicated values of $z$. }

\label{tsb}

\end{center} 
\end{figure}

The results for $R$ and $R^\prime$ following this procedure are illustrated in
Fig.\ \ref{tsb}.  

\begin{figure}
\begin{center}
\includegraphics[width=12cm]{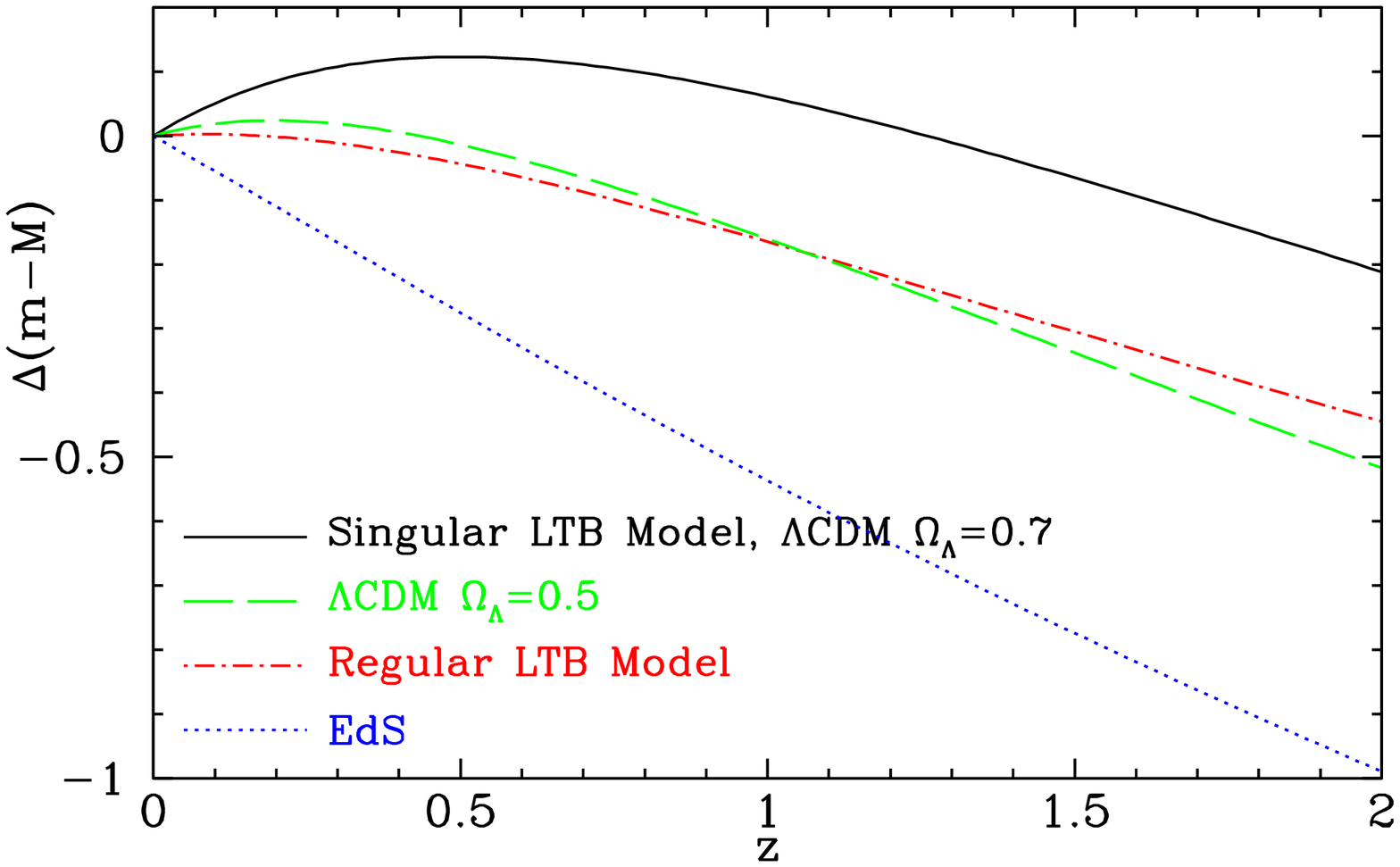}

\caption{The luminosity-distance vs.\ redshift for several cosmological models.
The regular LTB model is the LTB model with $\alpha(r)$ and $\beta(r)$ from
Sect.\ \ref{recon}, but with regular initial conditions. Also shown for
comparison is $\hat{d}_L(z)$ for a $\Lambda$CDM model with $\Omega_\Lambda=0.5$
instead of the fiducial model of this paper ($\Omega_\Lambda=0.7$). The singular
LTB solution is constructed to give the exact $\hat{d}_L(z)$ as for the fiducial
$\Lambda$CDM model.  Finally, in keeping with tradition, the result for the
Einstein-de Sitter model (spatially flat dust cosmology) is shown.  Here
$\Delta(m-M)$ is the difference in magnitudes between the indicated model and
the open, empty cosmological model. ($M$ is a nuisance parameter related to the
absolute magnitude of supernovae.)}

\label{dlz2}

\includegraphics[width=12cm]{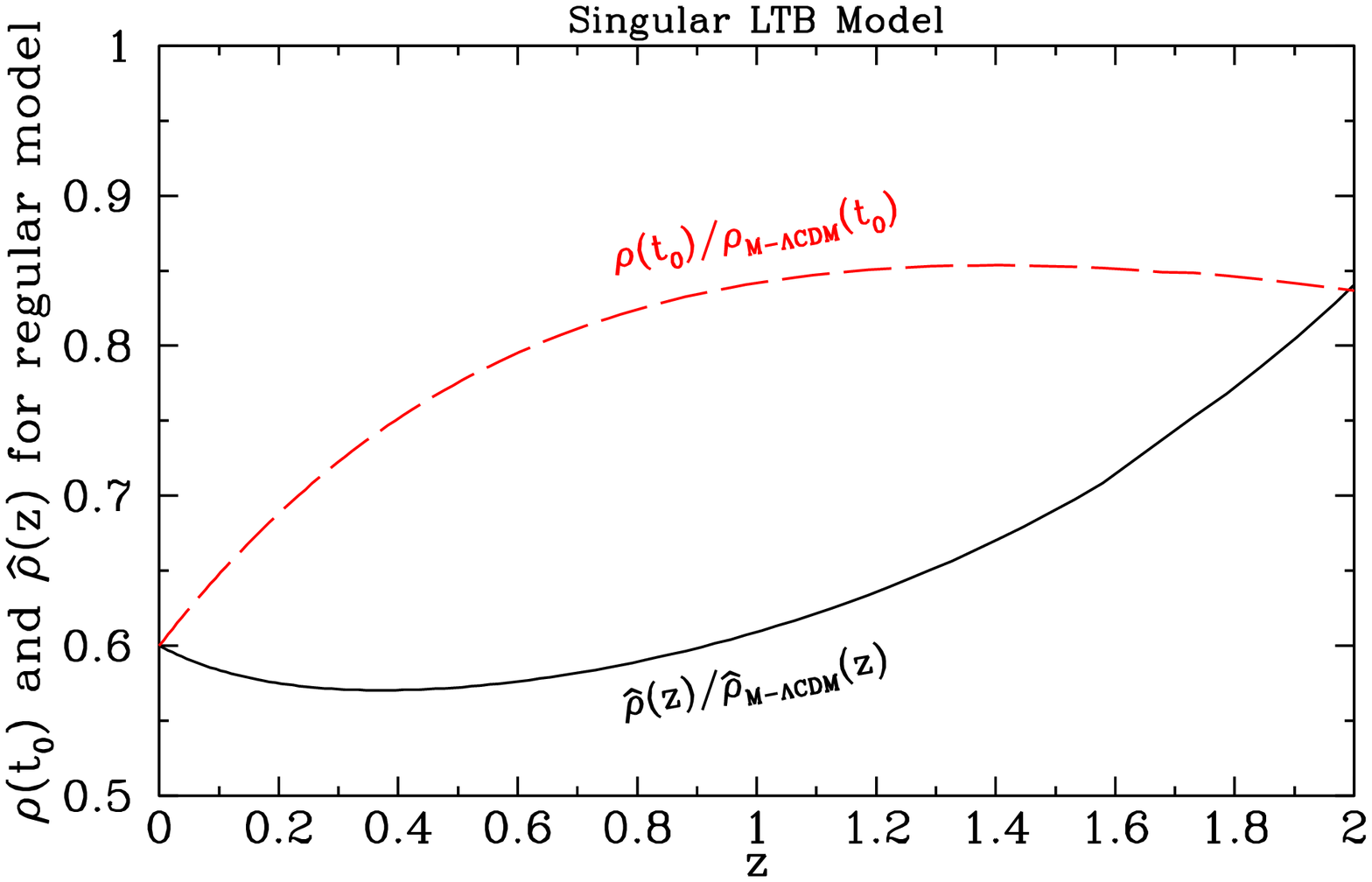}

\caption{The density as a function of redshift on the light cone (solid curve)
and at the present time (dashed curve) for the regular LTB model constructed in
Sec.\ \ref{recon}.  For the regular LTB model, the reference $\Lambda$CDM model 
has $\Omega_\Lambda=0.5$. }

\label{rhonbbbb}

\end{center} 
\end{figure}

Now consider the observational difference between the regular and singular
solutions.  In Fig.\ \ref{dlz2} we show the difference in $\hat{d}_L(z)$ for the
two solutions.  The regular solution is distinguishable from the singular
solution, but the singular solution closely resembles another $\Lambda$CDM
model, one with $\Omega_\Lambda=0.5$ rather than $\Omega_\Lambda=0.7$ of the
fiducial model, at least out to redshift two. In fact, out to redshift unity,
the difference is less than about 0.02 magnitudes.

Just as the luminosity distance for the regular cosmological solution differs
from the fiducial $\Lambda$CDM model, the density does as well.   This is
illustrated in Fig.\ \ref{rhonbbbb}.  Note that the reference $\Lambda$CDM model
for this figure is the model with $\Omega_\Lambda=0.5$, which well fits the
$\hat{d}_L(z)$ for the regular LTB model.  This figure well illustrates that the
density along the light cone compared to a fiducial model can decrease as a
function of $z$, while at the present time, it can increase as a function of
$r$.

It is beyond the purpose of this paper to see whether one can fiddle with
$\alpha(r)$ and $\beta(r)$ to find a regular solution that is a better fit to
the fiducial $\Lambda$CDM model.

The conclusion of this section is that if one attempts to construct an LTB model
that agrees exactly with $\hat{d}_L(z)$ and $\widehat{\rho}(z)$ of the fiducial
$\Lambda$CDM model, the resulting model will have singular initial conditions
for $R^\prime$. We have argued that this is to be expected.  We surmise that
this will be true in general, and is not specific to our choice of the fiducial
$\Lambda$CDM model.  We also demonstrated that one can find a regular solution
with the same values of $\alpha(r)$ and $\beta(r)$ that results in a
$\hat{d}_L(z)$ and $\widehat{\rho}(z)$ that approximates a $\Lambda$CDM model
close to the fiducial model at late time.  

In the next section we return to the singular models and examine whether they
can be described in terms of averaged quantities.

\section{Description in terms of an averaging procedure \label{average}}

One proposed route to understanding the role of inhomogeneities in mimicking
dark energy is by employing an averaging procedure.  Our calculations allow
us to study this proposal.  First, let us review the averaging procedure. 

A fundamental quantity in the analysis is the velocity gradient tensor, which is
defined as
\begin{equation}
\Theta^i_{\ j}=u^i_{\ ;j} = \frac{1}{2} h^{ik} \dot{h}_{kj},
\end{equation}
where $h^{ij}$ is the spatial metric and $u^\mu$ is the fluid four velocity.  
The tensor $\Theta^i_{\ j}$ represents the extrinsic curvature of the spatial
hypersurfaces orthogonal to the fluid flow.  It may be decomposed in terms of a
trace term and a traceless tensor as
\begin{equation}
\Theta^i_{\ j} = \Theta \, \delta^i_{\ j} + \sigma^i_{\ j} ,
\end{equation}
where $\Theta$ is the volume-expansion scalar.  The traceless tensor 
$\sigma^i_{\ j}$ is the shear.

The evolution equations for the expansion and the shear come from the
space-space components of Einstein's equations (see \textit{e.g.,} Ref.\
\cite{Matarrese:1995sb}).  Combining the expansion evolution equation with 
the energy constraint gives the \textit{Raychaudhuri equation},
\begin{equation}
\dot{\Theta} + \frac{1}{3} \Theta^2 + 2 \sigma^2 + 
\frac{1}{2}\kappa\rho =0. 
\end{equation}

From the Raychaudhuri equation it is straightforward to verify that local fluid
elements cannot undergo accelerated expansion.  (This point was emphasized by
Hirata and Seljak \cite{Hirata:2005ei}.) But, of course, as now appreciated,
that point is irrelevant.   Locally the expansion does not accelerate, but it is
incorrect to assume that acceleration does not occur when the fluid is
coarse-grained over a finite domain. The reason is trivial: the time derivative
of the average of $\Theta$ and the average of the time derivative of $\Theta$
are not the same because of the time dependence of the coarse-graining volume.

There has been a lot of work recently regarding the averaging procedure.  First,
we take the original proposal and average at a fixed time.  We will then average
over the light cone.

Let us denote the coarse-grained value of a quantity ${\cal F}$ by its
average over a spatial domain $D$:
\begin{equation}
\label{cg}
\langle {\cal F} \rangle_D = \frac{\int_D \sqrt{h} \,{\cal F}\, d^3\!x} 
{\int_D \sqrt{h}\,d^3\!x}.
\end{equation} 

Following the work of Buchert \cite{buchert1,buchert2}, we define a 
dimensionless scale factor
\begin{equation}
\label{dsf}
a_D(t) \equiv \left(\frac{V_D}{V_{D_0}}\right)^{1/3} ; \qquad 
V_D   = \int_D\, \sqrt{h}\,d^3\!x ,
\end{equation}
where $V_D$ is the volume of our coarse-graining domain (the subscript ``$0$'' 
denotes the present time).  In the LTB model
$\sqrt{h}=R^2(r,t_0)R^\prime(r,t_0)/\sqrt{1+\beta(r)}$.

The coarse-grained Hubble rate $H_D$ will be 
\begin{equation}
H_D = \frac{\dot{a}_D}{a_D}=\frac{1}{3}\langle\Theta\rangle_D .
\end{equation}
The smoothing procedure leads to the evolution equations for the coarse-grained
scale factor \cite{buchert1,buchert2}:
\begin{eqnarray}
\label{add}
\frac{\ddot{a}_D}{a_D}& = & -\frac{\kappa}{6}
\left(\rho_{\rm eff}+ 3 p_{\rm eff}\right) , \\
\label{ad}
\left(\frac{\dot{a}_D}{a_D}\right)^2 & = & \frac{\kappa}{3}\rho_{\rm eff} ,
\end{eqnarray}
where the effective energy density and pressure terms are ($^3\cal{R}$ is the 
spatial curvature)
\begin{eqnarray}
\label{rhoeff}
\rho_\textrm{eff} & = & \langle \rho\rangle_D-\frac{Q_D}{2\kappa}-
\frac{\langle ^3\cal{R}\rangle_D}{2\kappa}  \\
\label{peff}
p_\textrm{eff} & = & -\frac{Q_D}{2\kappa} + 
  \frac{\langle ^3\cal{R}\rangle_D}{6\kappa} ,
\end{eqnarray}
and we have introduced the {\it kinematical backreaction} 
\begin{equation}
\label{list}
Q_D=\frac{2}{3}\left(\langle\Theta^2\rangle_D-\langle\Theta\rangle_D^2\right)
-2\langle\sigma^2\rangle_D . 
\end{equation}
Finally, we can define an effective equation of state, $w_\textrm{eff}$, to be
$w_\textrm{eff} = p_\textrm{eff}/\rho_\textrm{eff}$.

Now for LTB models,
\begin{eqnarray}
\Theta & = & \Gamma^k_{0k} = 2 \frac{\dot{R}}{R} +
\frac{\dot{R^\prime}}{R^\prime} \nonumber \\
\sigma^2 & = & \frac{1}{2}\sum_k\left(\Gamma^k_{0k}\right)^2
- \frac{1}{6} \left( \sum_k\Gamma^k_{0k} \right)^2 
= \frac{1}{3} \left( \frac{\dot{R}}{R} - \frac{\dot{R^\prime}}{R^\prime} \right)
\nonumber \\
^3\cal{R} & = & -2 \frac{\beta}{R^2} - 2 \frac{\beta^\prime}{RR^\prime} .
\label{LTBM}
\end{eqnarray}

\begin{figure}
\begin{center}

\includegraphics[width=12cm]{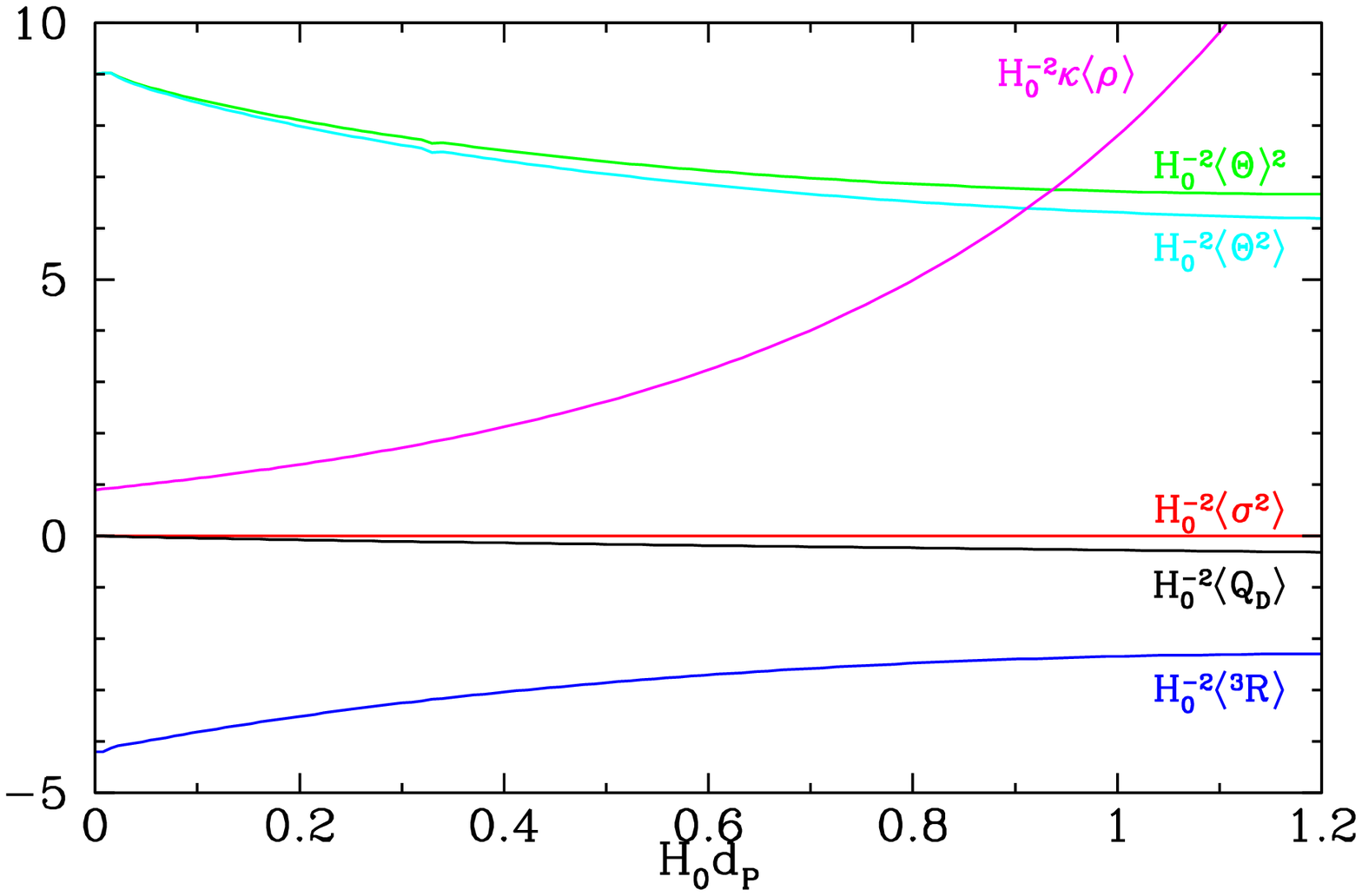}

\caption{The various terms that enter the calculation for $\rho_\textrm{eff}$
and $p_\textrm{eff}$ for the averaging at the present time as a function of
proper distance that sets the scale of the averaging volume. \label{avg0}}

\includegraphics[width=12cm]{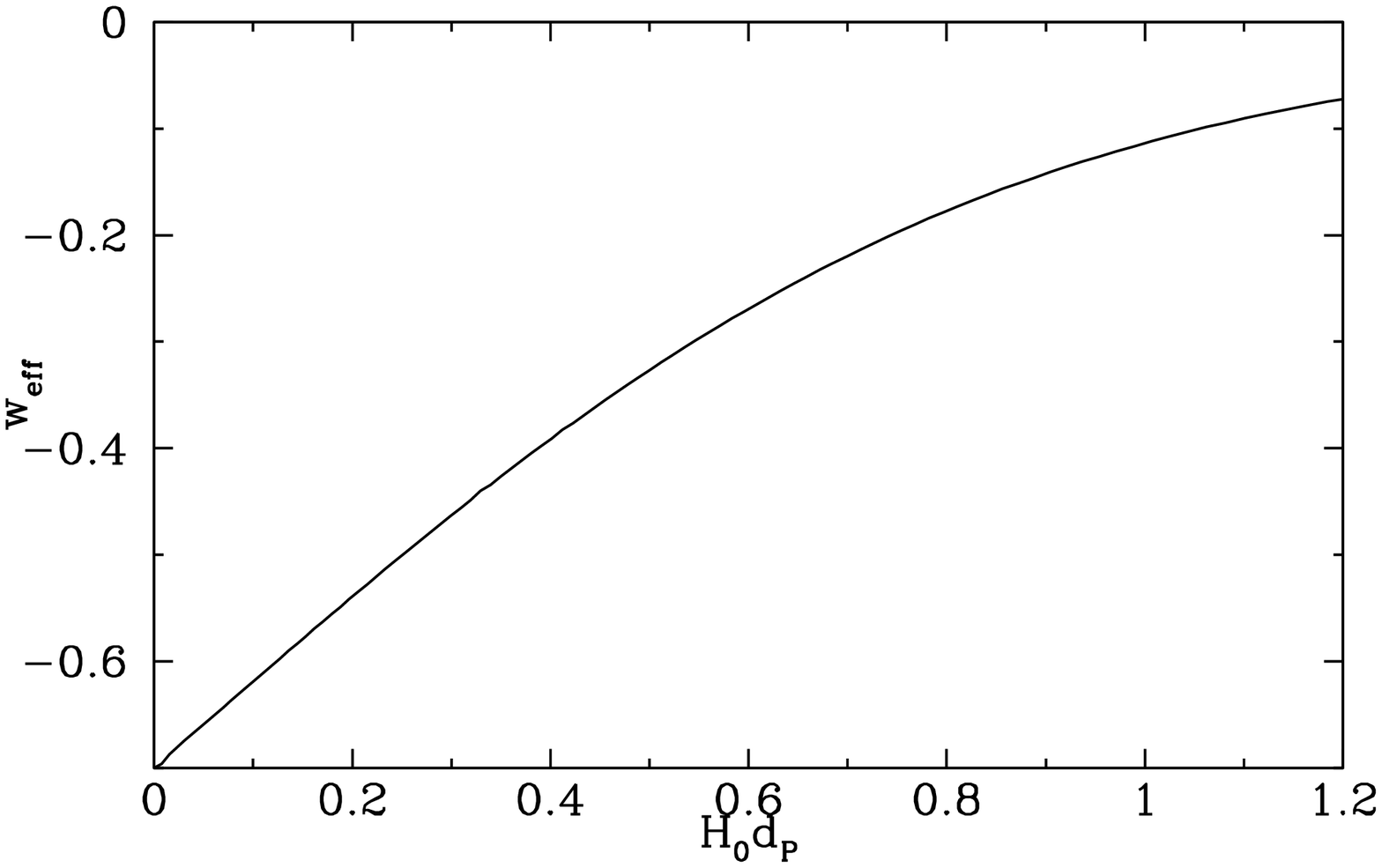}

\caption{The value of $w_\textrm{eff}$ as a function of proper distance that
sets the scale of the averaging volume for the averaging at the present time.
\label{w0}}

\end{center} 
\end{figure} 

\begin{figure}
\begin{center}

\includegraphics[width=12cm]{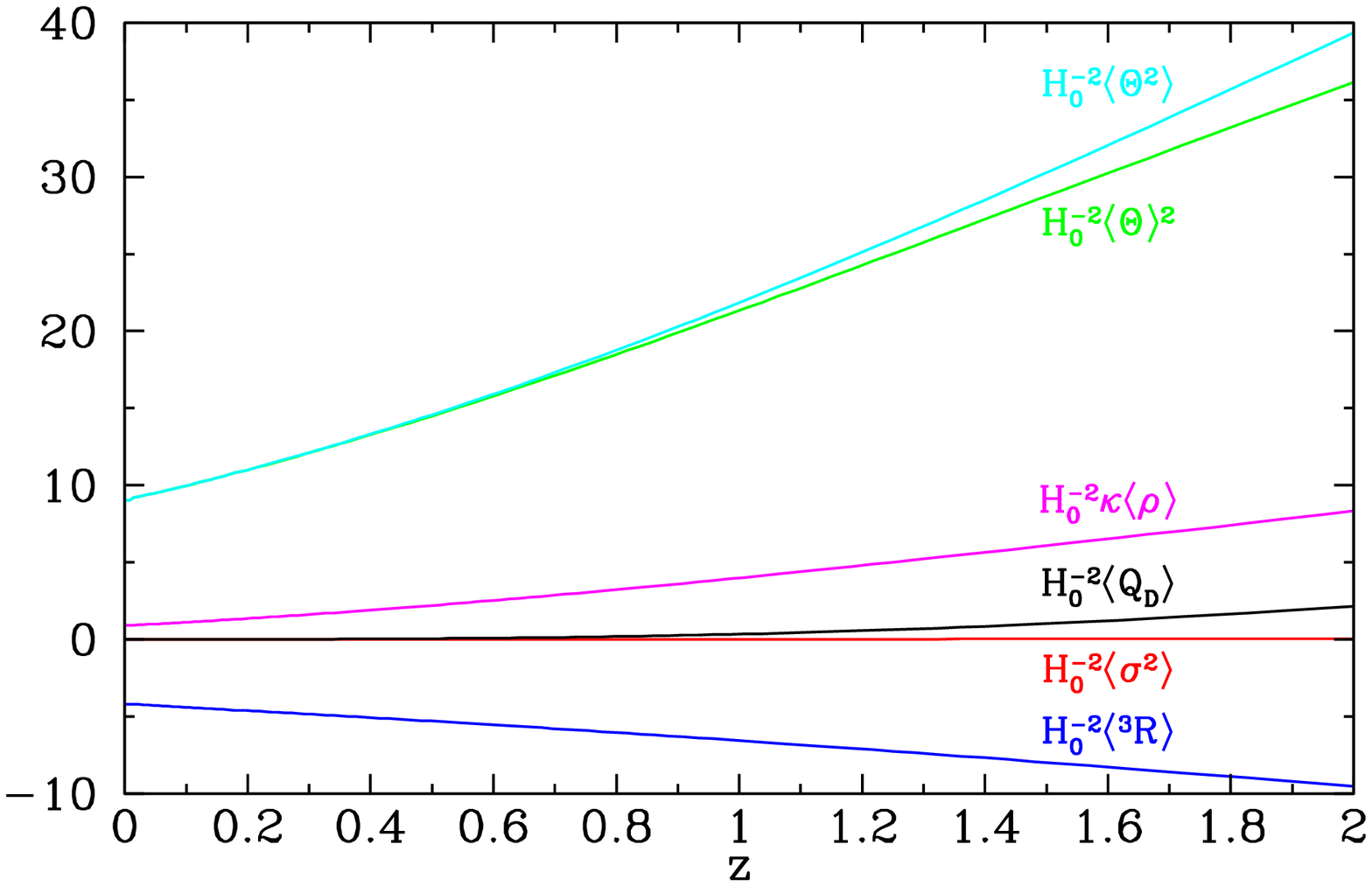}

\caption{The various terms that enter the calculation for $\rho_\textrm{eff}$
and $p_\textrm{eff}$ for light-cone averaging as a function of $z$.
\label{avghat}}

\includegraphics[width=12cm]{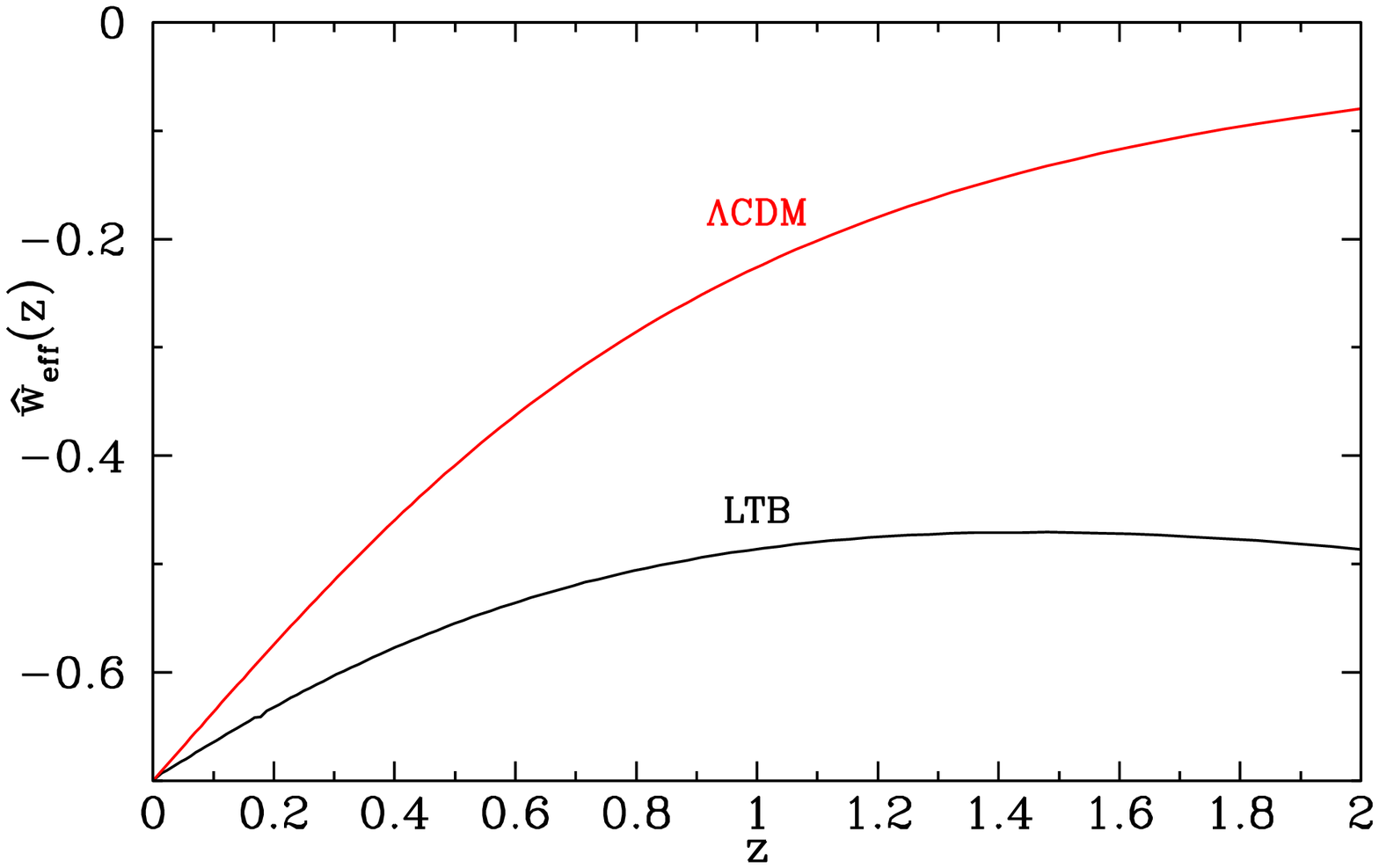}

\caption{The value of $w_\textrm{eff}$ as a function of $z$ for light-cone
averaging. \label{what}}

\end{center} 
\end{figure} 

The various terms that enter the calculation of $\rho_\textrm{eff}$ and
$p_\textrm{eff}$ are shown in Fig.\ \ref{avg0} as a function of the proper
distance that sets the scale of the averaging volume.  Note that the shear is
small, as is the kinematical back reaction since $\langle\Theta^2\rangle_D
\simeq \langle\Theta\rangle_D^2$. Shown in Fig.\ \ref{w0} is the effective value
of $w$, $w_\textrm{eff}=p_\textrm{eff}/\rho_\textrm{eff}$.  Note that if one
models the fiducial $\Lambda$CDM model as a single fluid and calculates $w$, the
value today is $w=-0.7$.

Another approach is light-cone averaging.  Here, what is of interest isn't the
average on a (acausal) constant-time hypersurface, but rather the average on a
hypersurface coinciding with the light cone.  One way to approach this is just
to substitute the ``hatted'' quantities in Eqs.\ (\ref{dsf}-\ref{LTBM}).  The
resulting terms that enter the calculation of the averaged quantities are shown
in Fig.\ \ref{avghat}, and the effective value of the equation of state
parameter is shown in Fig.\ \ref{what}.  Also shown in Fig.\ \ref{what} is the
equation of state parameter for the fiducial $\Lambda$CDM model if one would
model the total energy (including matter and $\Lambda$) as a single fluid.

There are two things we learn from the exercise of this section. First, neither
averaging on a constant-time hypersurface nor light-cone averaging is easy to
connect with the observations corresponding to parameters of the $\Lambda$CDM
model.  However, for both averaging prescriptions the effective
equation-of-state parameter is negative, indicating acceleration of some
averaged scale factor.  This may be a valuable check in more complicated
models.  It is interesting to note that the shear contribution to the evolution
of the averaged quantities is insignificant in both averaging prescriptions. 
This seems to be a counterexample to na\"{i}ve claims that if one constructs an
inhomogeneous universe to mimic the observations in $\Lambda$CDM one would
generate unacceptably large shear terms.

\section{Discussion and conclusions \label{conclusions}}

In the discussion of what we have learned in this investigation, it is useful
again to refer the reader to Fig.\ \ref{rhat} and to reiterate that all direct
cosmological information we have is limited to the light cone, illustrated by
the solid line in the figure.  Any statement about the Universe off of the light
cone is unsupported by direct observations, and hence requires a cosmological
model for the evolution of the Universe.  

An interesting lesson from the consideration of this paper is that if one is
just restricted to the two cosmological observables employed in this paper,
$\hat{d}_L(z)$ and $\widehat{\rho}(z)$, it is impossible to  determine
\textit{uniquely} a cosmological model \cite{Mustapha:1998jb,Celerier:2009sv}.
In our case, the fiducial $\Lambda$CDM model and the singular LTB model result
in identical observations for $\hat{d}_L(z)$ and $\widehat{\rho}(z)$.  That
result is valid even if one imagines perfect astronomical observations, since by
construction the two cosmological observables are degenerate in the two models.

\begin{figure}
\begin{center}
\includegraphics[width=12cm]{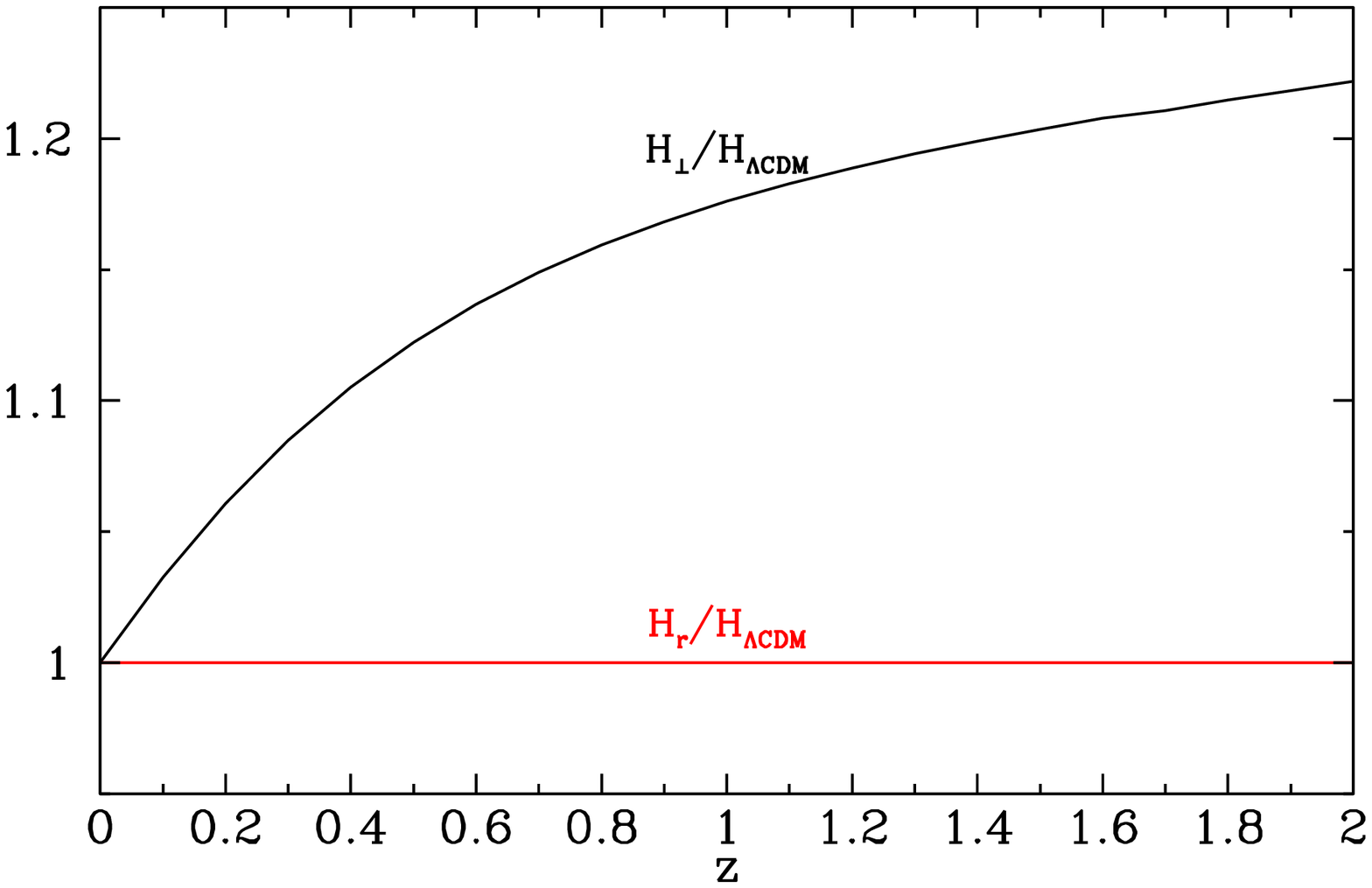}

\caption{The expansion rates as a function of redshift compared to the result
for the fiducial $\Lambda$CDM model. The expansion rates $H_r$ and $H_\perp$
were defined in Sect.\ \ref{LTB}.}

\label{h}

\end{center} 
\end{figure}

There are two ways to break the degeneracy, one observational and the other
theoretical.   One promising avenue to break the degeneracy with observational
data might be to take advantage of the fact that in LTB models there are two
different expansion rates, a radial expansion rate $H_r$, and an angular
expansion rate $H_\perp$ (they were defined in Sect.\ \ref{LTB}). Figure \ref{h}
shows the differences between the expansion rates in the singular model.  Note
that the fact that $H_r=H_{\Lambda CDM}$ is guaranteed by the input requirements
used to construct the model. Quartin and Amendola \cite{Quartin:2009xr} have
recently discussed using cosmic parallax and redshift drift to distinguish
between void models and dark energy. It is possible to pile on additional
observational constraints to break the degeneracy.  

While this can be done, it misses the important point that if one is restricted
to light-cone observations, it is unclear whether any number of observations
alone, even of arbitrary precision and with no systematic uncertainties, can
uniquely determine a cosmological model, even if it is possible to rule out any
LTB model.  An FLRW model is a zero-dimensional (dimensions of space) model, and
the LTB solution is a one-dimensional model. Intuition from other fields lead us
to believe that the range of solutions in two-dimensional, or fully
inhomogeneous three-dimensional models, will be richer.  So it is unclear
whether any number of cosmological observations, even to arbitrary accuracy, can
uniquely specify a cosmological model. 

This point was discussed by Kolb, Marra, and Matarrese (KMM)
\cite{Kolb:2009rp}.  In that work they introduce the concept of cosmological
background solutions.  This concept will be useful in elucidating several
points, so we briefly review some concepts from KMM.

A cosmological background solution is defined as a mean-field geometry of
suitably averaged Einstein equations, in which the average expansion is
described by a single scale factor. A cosmological background solution depends
upon the spatial curvature and the mass-energy content. Associated with the
mass-energy content is a stress-energy tensor that describes a fluid (or fluids)
with equation(s) of state that satisfy local energy conditions.

KMM then define several types of background solutions; of relevance here is the
\textit{Phenomenological Background Solution (PBS)}: a background solution that
effectively describes the observations, that is, the data on the past light cone
of an observer. The energy content and curvature of the PBS are not necessarily
related to the spatial averages of the energy content and curvature of the
observable universe. The equation of state of the stress-energy tensor of the
PBS need not satisfy any of the local energy conditions of the mass-energy
content of the universe. 

In this nomenclature, one might imagine that the true background solution is the
LTB solution, or more generally some inhomogeneous model, and the fiducial
$\Lambda$CDM solution is merely a phenomenological background solution. It is
unclear whether it is possible, in principle, to exclude this possibility just
on the basis of light-cone observations.

Observations alone may not be able to determine the true cosmological model, but
as remarked in the introduction, we expect the cosmological model to be part of
our deeper comprehension of the physical and astronomical facts; it must fit
into the larger framework of particle physics, general relativity, astrophysics,
and cosmology.

As an example of larger considerations, consider the theory of structure
formation.  In the fiducial $\Lambda$CDM model, structure grows in a background
that is close to homogeneous and isotropic, but with small seed perturbations
initially generated by inflation. Even in this zero-dimensional model, structure
formation is a complicated affair (at least in the nonlinear regime). Structure
formation in LTB models is an even more complicated affair
\cite{Biswas:2006ub,book}: each spherical shell evolves as an independent FLRW
model.  One would expect the evolution of structure formation in LTB models to
differ from the fiducial $\Lambda$CDM model. However, since structure formation
in a three-dimensional model is more complicated still, it is difficult to see
how structure formation might be conclusive.  The same argument might apply to
other cosmological tests like the integrated Sachs--Wolfe effect.

Now, we return to the argument that purports to prove that inhomogeneous models
cannot mimic the effects of dark energy (see, \textit{e.g.,} Refs.\
\cite{lamuros,Siegel:2005xu,Ishibashi:2005sj}).  The argument might be
paraphrased as follows: Perform an expansion of the FLRW metric to  Newtonian
order in potential and velocity but take into account the fully nonlinear
density inhomogeneities.  In the conformal Newtonian gauge, the metric will be
of the form
\begin{equation} 
ds^2 = a^2(\eta) [-(1 + 2 \psi) d\eta^2 + (1 - 2 \phi) d\bf{x}^2] ,
\end{equation} 
where $\eta$ is conformal time and $\phi \simeq \psi$.  The volume average of 
the equation then yields \cite{Siegel:2005xu}
\begin{equation} 
\left( \frac{\dot{a}}{a} \right)^2 = \frac{\kappa}{3} \langle\rho\rangle 
\left( 1 - 5W + 2K \right) , 
\end{equation}
where $W$ and $K$ are the Newtonian potential and kinetic energy (per unit 
mass),  
\begin{equation} 
\label{wandk}
W = \frac{1}{2} \langle (1 + \delta) \phi \rangle , \qquad 
K = \frac{1}{2} \langle (1 + \delta) v^2 \rangle .
\end{equation}
Here, $\delta = \delta \rho / \langle\rho\rangle$ is the density perturbation in
the matter rest frame and $v$ is the peculiar velocity.   The argument is that
observationally both $W$ and $K$ are small, hence, inhomogeneities cannot mimic
dark energy.

However the model constructed in Sec.\ \ref{recon} seems to belie this simple
and compelling argument.  If one interprets the observations of
$\widehat{\rho}(z)$ of the LTB model in terms of the fiducial $\Lambda$CDM
phenomenological background solution, one would conclude that the perturbations
are very small (\textit{viz.}\ zero), hence $W$ and $K$ zero, and
inhomogeneities could not mimic dark energy. But, at least for $\hat{d}_L(z)$
and $\widehat{\rho}(z)$, inhomogeneities do exactly that!

We point out that the very notion of peculiar velocities requires a background
solution (a norm) about which to define ``peculiarity'' \cite{Kolb:2008bn}. 
Also, Enqvist, Mattsson, and Rigopoulos \cite{Enqvist:2009hn} explicitly
demonstrate how spherically symmetric perturbations of a dust Universe can
lead to acceleration while the perturbations of the gravitational potential
remain small on all scales.  Perhaps the general lesson is that there are
highly non-Gaussian inhomogeneities in the late Universe, and the coherence of
structures causes small deviations in observables to sum to a large deviation.

We can venture a different way to frame the argument: If the observed density
fluctuations, $\delta\widehat{\rho}(z)$, originated from the growth of
perturbations in an FLRW cosmology, and the nonlinear perturbations on one scale
did not affect the growth of structure or the evolution of the background
solution on any other scale, and inhomogeneities do not show coherence or
non-Gaussianity, and $\phi\simeq\psi$, then one can say that the interpretation
of observations would yield a value of $W$ and $K$ that are too small to mimic
dark energy. 

This way of framing the argument makes manifest the fact that there are
dynamical assumptions that enter, and it cannot simply be a matter that small
peculiar velocities rule out the backreaction proposal.  If any of the dynamical
assumptions fail, the argument fails.

Another lesson that can be extracted from the exercise of this paper concerns
using averaging procedures to quantify the effect of inhomogeneities on
observables. This was the subject of the previous section. The conclusion was
that while the two averaging prescriptions discussed give an indication that the
LTB model would have observables that might be interpreted as arising from dark
energy, the effective equation of state parameters do not resemble that of the
fiducial $\Lambda$CDM model.  This might be because the averaging approach is
not very useful, or it might imply something quite troubling. Perhaps the
equation of state parameter $w$, which has received so much attention, has a
physical interpretation only in FLRW models, and in an inhomogeneous Universe,
where there is no natural time slicing, it is an unphysical construct.

This paper has developed and discussed issues related to the interpretation of
cosmological observables in constructing cosmological models and theories. 
Observations of $\hat{d}_L(z)$ and $\widehat{\rho}(z)$ do not, by and of
themselves, prove that there is dark energy or that the Universe is accelerating
in the usual sense; they do so only if one assumes that the Universe is
homogeneous and isotropic and the dynamics of the Universe are governed by
general relativity. This paper discussed aspects of the program to develop
understanding of how an inhomogeneous Universe might mimic dark energy.  While
consideration of the results of this investigation might lead one to conclude
that there are no arguments that conclusively exclude the backreaction proposal,
we are no closer to developing an inhomogeneous cosmology alternative to dark
energy.

One attempt to construct inhomogeneous cosmological models involves studying
``Swiss-cheese'' models comprised of low density voids embedded in an
Einstein--de Sitter background
\cite{Sugiura:1999fm,Kozaki:2002ka,Brouzakis:2006dj,Marra,Bolejko:2008xh,Vanderveld:2008vi,Brouzakis,Ghassemi:2009ug,Clifton:2009nv}. 
If the void regions are packed together the regions between them form a network
of high-density filamentary, or spaghetti-like, structures.  Further refinements
of this approach have included spherical mass concentrations referred to as
``meatballs'' \cite{Kainulainen:2009sx}.  A Swiss-cheese, meatball, and
spaghetti cosmological model may include ingredients that are desirable on their
own, or work well in combination with one other ingredient, but combining them
all may lead to a model that is not easily digestible. 

Whatever direction the study of inhomogeneous cosmologies might take,
consideration of the results of this investigation should be important.

\acknowledgments{It is a pleasure to thank Marie-N\"{o}elle C\'el\'erier,
Valerio Marra, Sabino Matarrese, Albert Stebbins, and Dan Chung for useful 
discussions.  This work was supported in part by the Department of Energy.}



\end{document}